\documentstyle[12pt,aasms4]{article}
\tighten 
\begin{document}

%

\newcommand{\lya}{Lyman~$\alpha$}
\newcommand{\lyb}{Lyman~$\beta$}
\newcommand{\za}{$z_{\rm abs}$}
\newcommand{\ze}{$z_{\rm em}$}
\newcommand{\cmtwo}{cm$^{-2}$}
\newcommand{\nhi}{$N$(H$^0$)}
\newcommand{\nzn}{$N$(Zn$^+$)}
\newcommand{\ncr}{$N$(Cr$^+$)}
\newcommand{\degpoint}{\mbox{$^\circ\mskip-7.0mu.\,$}}
\newcommand{\halpha}{\mbox{H$\alpha$}}
\newcommand{\hbeta}{\mbox{H$\beta$}}
\newcommand{\hgamma}{\mbox{H$\gamma$}}
\newcommand{\kms}{\,km~s$^{-1}$}      
\newcommand{\minpoint}{\mbox{$'\mskip-4.7mu.\mskip0.8mu$}}
\newcommand{\mv}{\mbox{$m_{_V}$}}
\newcommand{\Mv}{\mbox{$M_{_V}$}}
\newcommand{\peryr}{\mbox{$\>\rm yr^{-1}$}}
\newcommand{\secpoint}{\mbox{$''\mskip-7.6mu.\,$}}
\newcommand{\sqdeg}{\mbox{${\rm deg}^2$}}
\newcommand{\squig}{\sim\!\!}
\newcommand{\subsun}{\mbox{$_{\twelvesy\odot}$}}
\newcommand{\et}{et al.~}

\def\ltsima{$\; \buildrel < \over \sim \;$}
\def\simlt{\lower.5ex\hbox{\ltsima}}
\def\gtsima{$\; \buildrel > \over \sim \;$}
\def\simgt{\lower.5ex\hbox{\gtsima}}
\def\arcs{$''~$}
\def\arcm{$'~$}
\title{DUST IN HIGH REDSHIFT GALAXIES}
\author{\sc Max Pettini, David L. King}
\affil{Royal Greenwich Observatory, Madingley Road, Cambridge, CB3 0EZ, UK}
\affil{e-mail: pettini@ast.cam.ac.uk, king@ast.cam.ac.uk}
\author{\sc Linda J. Smith}
\affil{Department of Physics and Astronomy, University College
London,}
\affil{Gower Street, London WC1E 6BT, UK}
\affil{e-mail: ljs@star.ucl.ac.uk}
\author{\sc Richard W. Hunstead}
\affil{School of Physics, University of Sydney, NSW 2006, Australia}
\affil{e-mail: rwh@astrop.physics.usyd.edu.au}

\begin{abstract}

Measurements of Zn and Cr abundances in 18 damped \lya\ systems (DLAs)
at absorption redshifts
$z_{\rm abs} = 0.692 - 3.390$ (but mostly between $z \simeq 2$ and
3) show that metals and dust are much less abundant in high redshift
galaxies than in the Milky Way today. Typically, [Zn/H] $\simeq -1.2$;
as Zn tracks Fe closely in Galactic stars of all metallicities and is 
only lightly depleted onto interstellar grains, we
conclude that the overall degree of metal enrichment of damped \lya\
galaxies $\approx 13.5$~Gyr ago ($H_0 = 50$~km~s$^{-1}$~Mpc$^{-1}$,
$q_0 = 0.05$) was $\sim 1/15$ solar. 

Values of [Cr/Zn] span the range from $\simeq 0$ to $\simlt -0.65$ which
we interpret as evidence for selective depletion of Cr onto dust in
some DLAs. On average Cr and other refractory elements are depleted by
only a factor of $\approx 2$, significantly less than in local
interstellar clouds. We propose that this reflects an overall lower
abundance of dust---which may be related to the lower metallicities,
likely higher temperature of the ISM, and higher supernova rates in
these young galaxies---rather than an ``exotic''
composition of dust grains.

Combining a metallicity $Z_{\rm DLA} \simeq 1/15 Z_{\sun}$ with a 
dust-to-metals ratio $\approx 1/2$ of that in local interstellar 
clouds, we deduce that the ``typical'' dust-to-gas ratio in damped \lya\ 
galaxies is $\approx 1/30$ of the Milky Way
value. This amount of dust will introduce an extinction 
at 1500 \AA\ of only $A_{1500} \approx 0.1$ in the spectra of background QSOs.
Similarly, we expect little
reddening of the broad spectral energy distribution of the high-$z$ field
galaxies now being found routinely by deep imaging surveys. Even
such trace amounts of dust, however, can explain the weakness
of \lya\ emission from star-forming regions. We stress the approximate
nature of such general statements; in reality, the range of
metallicities and dust depletions encountered indicates that some
sight-lines through high-redshift galaxies may be essentially
dust-free, while others could suffer detectable extinction.

Finally, we show that, despite claims to the contrary, these
conclusions are {\it not} inconsistent with recent high resolution
observations of DLAs with the Keck telescope. We point out that the
star-formation histories of high-$z$ galaxies are not necessarily the
same as that of the Milky Way and that, if depletions of some 
elements onto dust are not taken into account correctly, it is possible to
misinterpret the clues to early nucleosynthesis provided by
non-solar element ratios.

\end{abstract}

\section{INTRODUCTION}

The question of whether high-redshift galaxies contain interstellar
dust has important implications for our interpretation of the
fast-growing body of data on the universe at $z > 2$. Dust may redden
the spectral energy distribution of galaxies sufficiently to mimic
the presence of an old stellar population (Steidel et al. 1996a).
Estimates of star-formation rates from the observed R-band
magnitudes---which at $z \simeq 3$ sample the rest-frame UV
continuum---are highly dependent on extinction corrections (e.g. 
Ellingson et al. 1996) and the transfer of \lya\ radiation is radically
affected even by trace amounts of dust (Charlot \& Fall 1993; Chen \&
Neufeld 1994). More generally, obscuration by dust in galaxies may
give us an increasingly biassed view of the universe with decreasing
redshift if the dust, as well as the metal, content of the universe
increases with time (Pei \& Fall 1995; Fall, Charlot, \& Pei 1996). 

Studies of damped \lya\ systems (DLAs) in QSO spectra 
offer several tests for the presence of dust.
Pei, Fall, \& Bechtold (1991)
found that QSOs with damped \lya\ systems at
$z \simeq 2-3$ tend on average to have slightly redder spectra 
than those of QSOs without DLAs; depending on the shape of the
extinction curve, the typical dust-to-gas ratio implied is between 
5\% and 20\% of that in the interstellar medium (ISM) of our
Galaxy today.

We have been pursuing a complementary approach which has the advantage
of yielding estimates of the dust-to-gas ratio in {\it individual}
DLAs, rather than an average for a given sample. The method, which is
described in \S2 below, is based on the comparison of the relative
gas-phase abundances of Zn and Cr, two elements which in the local ISM
have different dust depletion properties. The first results from our
survey of these two species were published in Pettini et al. (1994). 
In that paper we
reported that the typical dust-to-gas ratio in DLAs at $z \simeq 2$ is
approximately 10\% of the local value, although there is considerable
scatter.

Over the last three years we have continued our observations of Zn and
Cr in high-redshift galaxies; the full sample now includes 33 DLAs,
more than one third of the total number known. In this paper we
reconsider the evidence for dust in damped \lya\
systems using this enlarged data set, prompted in part by recent
claims that damped systems may be dust-free (Prochaska \& Wolfe 1996, 1997;
Lu, Sargent, \& Barlow 1997).\\

\section{METHOD}

As discussed by Pettini, Boksenberg, \& Hunstead (1990), in order to
use interstellar absorption lines to study the
chemical evolution of high-redshift galaxies it is necessary
to consider two effects: intrinsic departure from solar relative
abundances and selective depletion onto grains. The ratio of two
elements may be different from the value measured in the solar system
if the two elements are synthesized in processes which proceed at
different rates over the star formation history of a galaxy. An often
quoted example of this is the overabundance by about 0.5 dex of oxygen
and the $\alpha$-elements relative to iron in metal-poor ([Fe/H] $< -1$)
halo stars of the Milky Way\footnote{We use the normal notation where [X/Y] =
log(X/Y)$-$log(X/Y)$_{\sun}$}. O is produced primarily by massive stars
which evolve on much shorter timescales than the low-mass progenitors of
Type Ia supernovae thought to be the major source of 
Fe (e.g. Wheeler, Sneden, \& Truran 1989).
The presence of dust will superimpose on such nucleosynthetic
differences a depletion pattern reflecting the empirical
observation that some elements condense out of the gas phase into
solid form more readily than others (or alternatively, that some
elements are not as readily returned to the gas as others when
interstellar grains are destroyed in interstellar shocks).

In principle it should be possible to disentangle these two effects,
guided by existing information on stellar abundances
as a function of metallicity (e.g. Wheeler et al. 1989) and 
on the way element depletion varies in interstellar clouds in our
Galaxy (e.g. Savage \& Sembach 1996). It should be borne in mind,
however, that: (i) the star-formation histories of the galaxies giving rise
to damped \lya\ systems may differ from that of the
Milky Way at the same look-back time; and (ii) the composition of
interstellar dust may also be different in these young galaxies.

On the basis of these considerations Pettini et al. (1990) proposed
that Zn and Cr are a pair of elements well suited to measuring both
metal enrichment and dust depletion in damped \lya\ galaxies.
Observations summarised by Sneden, Gratton, \& Crocker (1991) show
that in Galactic stars [Zn/Fe] $= +0.04$, with little scatter ($\sigma =
0.1$~dex) over the range [Fe/H] $= -2.9$ to $-0.2$\,. 
In Large Magellanic Cloud stars [Zn/Fe] $\approx -0.2$
(Russell \& Dopita 1992). 
Evidently, the
productions of Fe and Zn are closely linked and their relative
abundances are, at least to first order, independent of metallicity, 
past star-formation history,
and galaxy mass. While it is true that the
reasons for such tight correspondence are not yet understood, this is
more likely to be an indication of the inadequacies of current
theoretical supernova yields rather than an indictment of the data
(Malaney \& Chaboyer 1996).

The production of Cr also follows closely that of Fe, since 
[Cr/Fe] $\simeq 0$ from [Fe/H] = 0 to $\simeq -2$\,. 
However, two recent surveys of extremely metal-poor stars by
McWilliam et al. (1995) and Ryan, Norris, \& Beers (1996) both show
that for [Fe/H] $\simlt -2.5$ [Cr/Fe] {\it decreases} significantly 
(and with considerable scatter). Apparently, the 
yield of Cr is reduced in the very low abundance
regime, possibly reflecting a shift to higher mass numbers 
in iron-peak nucleosynthesis (McWilliam et al. 1995).

In the interstellar medium of the Milky Way Zn and Cr have very different
dust depletion properties (see Fig. 2 of Savage \& Sembach 1996). Zn is at
most only lightly depleted and is present in near-solar proportions
in warm disk clouds and along sight-lines where the molecular fraction
is low. Cr, on the other hand, is among the most depleted elements;
typically between $\sim$90\% and $\sim$99\% of the Cr is in solid form.

Based on what is known on the relative abundances of Zn and Cr in
stars and the ISM of our Galaxy, we conclude that:

1) The Zn/H ratio in damped \lya\ systems is a measure of the degree
of metal enrichment, analogous to the [Fe/H] scale on which galactic
chemical evolution is normally reckoned; and

2) For [Zn/H] $\simgt -2$, departures from the solar value of the
Cr/Zn ratio are most naturally interpreted as being due to depletion
of Cr onto grains. By inference, the Cr/Zn ratio can be used to test
for the presence of interstellar dust in high redshift galaxies.\\

\section{OBSERVATIONS AND RESULTS}

The full data set consists of intermediate dispersion spectra of 33
damped \lya\ systems covering the Zn~II $\lambda\lambda 2026, 2062$ and
Cr~II $\lambda\lambda 2056, 2062, 2066$ resonance lines at absorption
redshifts ranging from $z_{\rm abs}$ = 0.692 to 3.390; 25 systems were
observed by us, 8 others are from the literature.
This sample constitutes a significant fraction of the total number
of damped systems known: the latest compilation by Wolfe et al. (1995)
lists 80 confirmed DLAs.
Our observations used the William Herschel telescope
on La Palma, Canary Islands, and the Anglo-Australian telescope
at Siding Spring Observatory, Australia.
The spectra are presented in separate papers (Pettini et al. 1994,1997),
together with details
of the data acquisition and reduction.

In Table 1 we have collected the measurements of relevance to the
determination of the dust content of DLAs, including absorption
redshift, neutral hydrogen column density $N$(H~I), [Zn/H], and [Cr/Zn]. 
In approximately half of the cases the Zn~II and Cr~II lines are below
the detection limit of our data (typically
$W_0$(3$\sigma$) = 25~m\AA, where $W_0$(3$\sigma$) is the
3$\sigma$ limit for the rest-frame equivalent width of an unresolved
absorption line), and 
[Cr/Zn] is therefore undetermined.
In total, we have been able to measure [Cr/Zn] (or place a useful
limit on this ratio) for 18 DLAs.

All the abundances listed in Table 1 were derived using the
experimental determinations of the $f$-values of the 
Zn~II and Cr~II multiplets
by Bergeson \& Lawler (1993) and the solar system abundances of Zn and
Cr from the compilation by Anders \& 
Grevesse (1989)\footnote{log~(Zn/H)$_{\sun} = -7.35$;
log~(Cr/H)$_{\sun} = -6.32$.}.
Consequently, for damped systems in common, 
the values of [Zn/H] and [Cr/Zn]
in Table 1 differ by $-0.148$ and $+0.276$ respectively
from those listed in Table 3 of Pettini et al. (1994)
which were based on earlier, 
theoretical estimates of the $f$-values and slightly
different solar abundances. The change, particularly to
[Cr/Zn], is significant and emphasises the importance of
accurate measurements of the atomic
parameters of these key transitions.

In Figure 1 the values of [Cr/Zn] measured in 18 damped \lya\ systems
(filled circles) are plotted against the metallicity [Zn/H]. 
The open circles are values typical of warm interstellar
clouds in the disk and halo of the Milky Way from the recent review by
Savage \& Sembach (1996). For halo clouds S was taken as a proxy for
Zn since few observations of the latter are available; this is a
reasonable assumption given that both S and Zn are undepleted in warm
interstellar clouds (Savage \& Sembach 1996).
We have also reproduced in the Figure 
the variation of [Cr/Fe] with [Fe/H] in Galactic
stars; the dotted lines 
correspond to the upper and lower
quartiles of the large sample of stars considered by Ryan et al. (1996).  

Figure 1 shows that there is a spread in the values of [Cr/Zn] in
damped \lya\ systems. In some cases [Cr/Zn] is in good agreement with
the relative abundances measured in stars which, over the range of
metallicities considered here, do not differ much from the
solar ratio. In many damped systems, however, [Cr/Zn] is 
lower than solar and comparable to values seen
in local halo clouds. 
Inspection of Figure 1 and column (5) of Table 1 
suggests that the range of [Cr/Zn] encountered is larger 
than the random scatter expected from measurement errors;
the largest and smallest value in the sample differ by more than 
$3.5 \sigma$. 

A straightforward average of our measurements 
of $N$(Cr$^+$)/$N$(Zn$^+$) yields a
mean $\langle$[Cr/Zn]$\rangle = -0.3^{+0.15}_{-0.2}$~($1 \sigma$ limits).
We interpret this result as an indication that in damped \lya\ systems
about half of the Cr is typically locked up in grains.\\

\section{DUST-TO-GAS RATIO AND THE REDDENING OF HIGH REDSHIFT GALAXIES}
\subsection{Depletion of Refractory Elements}

As already noted by Pettini et al. (1994), the degree of Cr depletion 
in damped \lya\ systems
is lower than that normally observed 
in interstellar clouds in the solar vicinity. 
However, it does not necessarily follow that the composition of
interstellar dust in these high-redshift galaxies is 
{\it qualitatively} different
from that in the local ISM, and is peculiarly deficient in Cr.
Generally, other refractory elements also show reduced
depletions in DLAs (e.g. Meyer \& Roth 1990); 
the most extensive data set available is for Fe
and indeed [Cr/Fe] $\simeq 0$ in 9 DLAs (Lu et al. 1997).

Of course we cannot exclude the possibility that there may be real differences
in the detailed composition of interstellar dust grains in
damped \lya\ systems as compared with the ISM of the Milky Way;
indeed, this may be an interesting topic to be addressed in future
with more precise determinations of the relative abundances of
different grain constituents than are available at present.
However, the most straightforward interpretation of the existing 
data is that, to a first approximation, {\it most} refractory elements are only
half-depleted in the ISM of the high-$z$ galaxies giving rise to
damped \lya\ systems. Another way of expressing this result
is that the {\it dust-to-metals} ratio in the DLAs 
is approximately half of that in the 
Galaxy.
 
While a detailed treatment is beyond the scope of this paper, 
this finding may not be difficult to explain. There are at least three
reasons why we may expect a lower overall level of heavy element
depletion:

1) Damped systems are generally metal-poor and the efficiency with which
refractory elements condense into solid form
may well vary non-linearly with the overall abundance of the grain
constituents. 

2) Grain destruction by interstellar shocks is likely to
be more effective at higher temperatures, as evidenced by the 
lower depletions in local warm interstellar clouds, compared with cooler
regions of the ISM (Savage \& Sembach 1996). At the low metallicities
of the damped \lya\ galaxies, it is likely that there is no cool
phase of the ISM (e.g. Hartquist \& Dyson 1984).

3) It has been suggested that the reduced depletions in the halo,
which for Cr approach the values we measure in DLAs (see Figure 1),
result from more frequent grain processing through
supernova-induced shocks (Savage \& Sembach 1996 and references
therein). With typical star-formation rates of
8.5~$M_{\odot}$~yr$^{-1}$ (Steidel et al. 1996a) 
the supernova rate in galaxies producing DLAs
is likely to be one order of magnitude higher than that of 
the Milky Way today.

\subsection{Dust-to-Gas Ratio}

Pettini et al. (1997) derive a column density-weighted mean
metallicity $\langle$[Zn/H]$\rangle = -1.15$ for the present sample of 33 
DLAs (individual values of [Zn/H] range over nearly $\pm 1$ dex 
about this mean value).
Consequently, the heavy elements which make up interstellar dust are,
on average, only $\sim 1/15$ as abundant as in the present-day ISM. 
If we further assume that
half of the grain constituents are in the gas-phase, as suggested by
our Cr and Zn observations, we conclude that the ``typical'' dust-to-gas
ratio in damped \lya\ systems is $\approx 1/30$ of that found in the
interstellar medium of our Galaxy.
This fraction may be an underestimate by a factor of $\approx 2$ if O,
which accounts for about half of the mass in grains, is {\it
overabundant} by a factor of $\sim 3$ relative to Zn, as is the
case in Galactic stars with [Zn/H] $= -1.15$\,. 

This estimate of the dust-to-gas ratio is 3 times lower than that
derived by Pettini et al. (1994). The change is {\it not} due to the
increased sample of DLAs, but mostly stems from the new $f$-values for
the Zn~II and Cr~II transitions adopted in the present analysis, as
explained in \S 3 above. Given the uncertainties and the dispersion
among different damped systems, we consider the new value of the ``typical''
dust-to-gas ratio still to be consistent with the range of
values (between $\approx 5$ and $\approx 20$ times lower than the local ISM) 
deduced by Pei et al. (1991) on the basis of the mild reddening they found in
QSOs with damped \lya\ systems.

\subsection{Reddening of the Spectra of QSOs and High-$z$ Galaxies}

In the disk of our Galaxy, 
$\langle N$(H~I)$\rangle / \langle A_V \rangle = 1.5 \times
10^{21}$~cm$^{-2}$~mag$^{-1}$ (Diplas \& Savage 1994), where $A_V$ is the
extinction (in magnitudes) in the $V$ band. For the typical damped \lya\ system
with neutral hydrogen column density $N$(H~I) = $1 \times 10^{21}$~cm$^{-2}$,
we therefore expect a trifling $A_V \simeq 0.02$~mag in the rest-frame 
$V$-band.  Of more interest is the
far-UV extinction, since this is the spectral region observed at optical
wavelengths at the redshifts of interest here ($z = 2 - 3$). Adopting the SMC
extinction curve (Bouchet et al. 1985)---which may be the appropriate one to
use at the low metallicities of most DLAs---we calculate that a damped
\lya\ system will typically introduce an extinction at 1500 \AA\ 
of $A_{1500} \simeq 0.1$~mag in the spectrum of a background QSO.
Such a small degree of obscuration is consistent
with the models of Pei \& Fall (1995)
and Fall, Charlot, \& Pei (1996) which assume the same 
dust-to-metals ratio as found here. In these models dust in galaxies
at $z \simeq 1$ introduces a significant bias in magnitude-limited QSO samples, 
but at $z = 2 - 3$ the effect is small.

The field galaxies at $z \simgt 3$ now being identified routinely by
means of deep Lyman limit imaging (Steidel et al. 1996a,b) exhibit
absorption spectra very similar to those of damped \lya\ systems, as was indeed
expected. In estimating the reddening of the stellar spectral energy
distribution, it is more appropriate to apply the
extinction law for starburst galaxies 
determined by Calzetti, Kinney, \& Storchi-Bergmann (1994).
In this case, $N$(H~I) = $1 \times 10^{21}$~cm$^{-2}$ would produce
$A_{1500} \simeq 0.05$~mag. 
Thus, {\it if} the dust-to-gas ratio of damped \lya\ systems
applies to the star-forming regions found by deep imaging, 
the spectral energy distributions of these objects are not 
significantly reddened by dust, irrespective of the details of the 
extinction curve.  However, even such a small amount of dust
would result in considerable quenching of \lya\ emission, given the 
large optical depths involved (e.g. Hartmann et al. 1988).
 
While the above considerations are of some interest, it is important
not to lose sight of the approximate nature of our estimate of the
abundance of dust. Given the range of values of [Cr/Zn] in
Figure 1 and the spread by $\sim 2$ orders of magnitude in [Zn/H] at
$z \simeq 2-3$ found by Pettini et al. (1994, 1997), the concept of a
``typical'' dust-to-gas ratio is questionable. Rather, the
observations suggest that some sight-lines through high-redshift
galaxies may be essentially dust-free, while others could suffer
detectable extinction. 

Finally, a dust-to-gas ratio 30 times lower than that of the Milky Way
implies that a $L^{\ast}$ galaxy with a mass of
$\sim 10^{11} M_{\sun}$ will contain $\sim 5 \times 10^7 M_{\sun}$
of dust. At $z = 2 - 3$, thermal emission from such quantities of dust
may be detectable between 850 and 450 $\mu$m with SCUBA, the new 
sub-millimeter array camera on the James Clerk Maxwell telescope on Mauna 
Kea, {\it if} the dust temperature is greater than 60~K.\\

\section{CONFLICT WITH KECK OBSERVATIONS?}

Recent analyses of echelle observations of damped \lya\ systems with
the HIRES spectrograph on the Keck telescope (Prochaska \& Wolfe 1996,
1997; Lu et al. 1997) have concluded that there is {\it no} evidence 
for dust depletion in these galaxies and have questioned the validity
of the method outlined here---based on the [Cr/Zn] ratio---as a test
for the presence of dust at high redshift. As a logical extension of
this reasoning, Lu et al. further suggest that the well-documented
lack of variation of [Zn/Fe] with [Fe/H] in Galactic stars may be
misleading.

It is important to realise at the outset that these authors'
conclusions are {\it not} based on improved measures of the Zn and Cr
abundances made possible by the powerful combination of HIRES and the
Keck telescope; in all cases in common there is good agreement with
the 4-m observations reported by Pettini et al. (1994, 1997). 

Rather, there are significant differences in the interpretation of the
common body of data. As we shall now show, the observations by
Prochaska \& Wolfe and by Lu et al. are in fact entirely consistent
with the overall picture proposed here, whereby: (i) dust (and metals) are
much less abundant in high-redshift galaxies than in the local ISM;
(ii) the ``typical'' dust-to-gas ratio is $\approx 1/30$ of that of
diffuse interstellar clouds in the Milky Way, but with considerable
spread between different sight-lines; (iii) no exotic composition of
interstellar grains is required, only an overall reduction in the
proportion of heavy elements locked up in solid form.

Before proceeding further, it is worth considering whether this is a
useful discussion; after all, could it not be argued that a dust
abundance as low as 1/30 of today's is, at least to a first
approximation, equivalent to no dust at all?  In reality, the qualitative
difference between these two possibilities has important consequences:

A) If all damped \lya\ systems were dust-free, as has been proposed,
this would be a major difference between galaxies at high redshift and
those in the nearby universe, where dust is a ubiquitous component of
the ISM.

B) Even small amounts of dust have a profound effect on the transfer
of \lya\ radiation, and will alter to some extent our view of the
distant universe if they give rise to a redshift dependent bias, as
emphasised by Fall and collaborators (see references cited above).

C) Even greatly reduced depletions of grain constituents may alter
relative element abundances by factors which are comparable to those
resulting from nucleosynthesis effects. Intrinsic departures from
solar ratios seen in Galactic stars of different metallicities amount
in many cases to factors of less than $\sim 3$; without taking into
account properly the effects of dust depletion, we may be misled in
our attempts to interpret the clues provided by non-solar element ratios.

The assertion by Prochaska \& Wolfe and by Lu et al. that the pattern
of element abundances seen in high-$z$ damped \lya\ systems is {\it
inconsistent} with dust depletions is based on two main lines of
evidence: 

1) There is no obvious correlation between degree of depletion of an
element and its condensation temperature, whereas such a trend is seen in
local interstellar clouds; and

2) Some element ratios exhibit departures from solar values
which are consistent with nucleosynthesis by Type II supernovae and
apparently admit no additional contribution from selective grain
depletions.

We now discuss each point in turn.

\subsection{Lack of a Correlation between Depletion and Condensation 
Temperature}

Comparisons with the local pattern of element depletions are normally
made with reference to the line of sight to $\zeta$~Oph; the large
depletions of refractory elements in the cool cloud in front of this
star indeed highlight a trend of decreasing gas-phase abundance with
increasing condensation temperature---see Figure 5 of Savage \&
Sembach (1996). So, for example, Ti is more depleted than Fe and Cr
which in turn are more depleted than Mn; the measured gas-phase
abundances [Ti/H] $= -3.02$, [Fe/H] $= -2.27$, [Cr/H] $= -2.28$, and [Mn/H]
$= -1.45$ imply that $\approx 99.9$\% of the Ti, $\approx 99$\% of the Fe
and Cr, and $\approx$96\% of the Mn are in solid form.

However, what has been overlooked is the fact that, at the reduced
dust depletions prevailing in damped \lya\ systems, such a pattern
would be largely washed out. Consider what happens if half of the
interstellar grains are destroyed, as we have proposed is the case in
the ``typical'' DLA. Then the gas-phase abundances of Ti, Fe, Cr and
Mn are {\it all} increased to $\approx$50\% and all 4 elements would
be found to be depleted by approximately the same amount to within a
few percent. (Such a dilution of the pattern of relative depletions can
be seen directly in warm interstellar clouds in the halo of the Milky
Way, which show lower overall depletions than cool disk clouds---see
Figure 6 of Savage \& Sembach 1996). 

Superimposed on this uniform degree of depletion we may then find
intrinsic departures from solar ratios resulting from differences in 
the nucleosynthetic origin of some elements. Thus [Mn/Fe] $\simeq
-0.3$ in DLAs with [Fe/H] $\simeq -1.0$ to $-1.6$ (Lu et al. 1997), as
is the case in Galactic stars of this metallicity (Ryan et al. 1997).

The fact that refractory elements exhibit reduced depletions in DLAs
has been apparent from the earliest studies of this kind (e.g. Meyer,
Welty, \& York 1989; Pettini et al. 1990) and is indeed evident in the
Keck data; the mistake is in assuming that the same relative gas-phase
abundances which result when nearly 100\% of these elements are locked
up in grains, and only tiny fractions remain in gaseous form, are
maintained when the overall degree of depletion is 
reduced to only $\approx 1/2$. 

Lu et al. have proposed that the [Mn/Fe] ratio can be used to
discriminate between dust depletion and intrinsic differences due to
nucleosynthesis because Mn is less depleted than Fe in cool
interstellar clouds and more deficient than Fe in metal-poor stars. In
view of the discussion above, however, we now see that this reasoning
no longer holds at low overall levels of depletion. Rather, the fact
that the [Mn/Fe] ratio responds to {\it both} effects makes it less
suitable for disentangling one from the other. A more straightforward
test for the presence of dust is provided by the ratios of elements
which are {\it not} sensitive to the nucleosynthetic history of the
gas, such as Cr, Ni, and Fe relative to Zn. Similarly, clues to the
nucleosynthesis of damped \lya\ systems are best deduced from elements
which are not easily depleted onto grains but have relative abundances
which do vary with metallicity, such as Zn and S.

\subsection{Nucleosynthesis by Type II Supernovae as the Sole Reason for 
Non-Solar Element Ratios}

We now consider the second point. Lu et al. argue that the
[N/O] and [S/Fe] ratios are inconsistent with dust depletions because they
exhibit the same values as in Galactic metal-poor stars.  
But, as dicussed by 
Pettini, Lipman \& Hunstead 1995 and in more detail by Lipman (1995),
the [N/O] ratio is more sensitive than most to the past history of
star formation in a galaxy.
For this reason, and given the uncertainties in the relative importance 
of primary and secondary production of N at low metallicities, 
[N/O] is not a useful tracer of dust. 

Turning to [S/Fe], Lu et al. find [S/Fe] $\simeq +0.4$, in line with the
well-known overabundance of the $\alpha$-elements at low
metallicities; depletion of Fe onto grains, if present, would have
been expected to increase this ratio further. The problem here is that
S has been measured in only {\it three} DLAs in total, two of which
(Q0000$-$263 and Q2348$-$148) are among the most metal-deficient
known, with [S/H] $\simeq -2$\, (see Figure 23 and Table 16 of Lu et
al.). Zn has not been detected in either of these two
systems, so we cannot test directly for the presence of dust. It
is quite conceivable (see Figure 1) that these extremely metal-poor
systems are indeed dust-free. However, on the basis of the large dispersion in
[Cr/Zn] evident in our sample of 18 DLAs, we see little justification
for extending this conclusion to {\it all} other damped \lya\ systems.

In Galactic stars [S/Zn] $\simeq +0.5$ when [Zn/H] = [Fe/H] $\simlt
-1$\,. If future observations were to show that [S/Zn] $\simeq 0$ in
DLAs---as is the case in the {\it one} system where both species have
been measured to date, at $z_{\rm abs} = 2.8110$ in Q0528$-$250---would
this be conclusive proof that Zn is in some way ``anomalous'' and, by
inference, that DLAs contain no dust, as reasoned by Lu et al.? Not
necessarily. An alternative explanation is that the star-formation
history of some high-$z$ galaxies may be different from that of the
Milky Way. The overabundance of the $\alpha$-elements in metal-poor
halo stars is thought to be the result of the time lag between Type II
and Type Ia supernovae in  {\it continuous}
star-formation models.
When star formation proceeds in bursts, however, different values of
[$\alpha$/Fe] can ensue, as emphasized by Gilmore \& Wyse (1991).
In the Magellanic Clouds, for example, O is {\it less} abundant than
Fe (Russell \& Dopita 1992). As Kennicutt (1995) has shown, galaxies
in the Local Group have undergone widely different star-formation
histories; it is likely that this is also the case for
the galaxies giving rise to damped \lya\ systems.

In conclusion, the range of values of the [Cr/Zn] ratio we have
measured in 18 damped \lya\ systems has led us to conclude that dust
is present in at least some high-redshift galaxies, although always in
lower concentrations than in the Milky Way ISM. We find no conflict
between this result and recently published Keck observations. Unless
the depletions of refractory elements are properly taken into account,
it will be difficult to use relative element abundances to unravel the
nucleosynthetic history of galaxies at high redshifts.

\acknowledgements
It is a pleasure to acknowledge the continuing support of this work
by the WHT and AAT time allocation
committees, and the La Palma service observations scheme.
We thank Sean Ryan for providing us with
his compilation of Cr and Fe abundances in metal-poor stars, 
Mike Fall for helpful comments on an early version of the paper, and 
Alec Boksenberg for his encouragement to publish this work.
Andrew Blain kindly provided estimates of the sensitivity of SCUBA for 
detecting dust at high-redshift.
R.W.H. gratefully acknowledges financial assistance 
by the Australian Research Council.

\begin{deluxetable}{lllllcc}
\tablewidth{0pc}
\tablecaption{Zn and Cr Abundances in DLAs}
\tablehead{
\colhead{Object} & \colhead{$z_{\rm abs}$} & \colhead{log~$N$(H~I)\tablenotemark{a}} & \colhead{[Zn/H]} & \colhead{[Cr/Zn]} & \colhead{Ref.}  }
\startdata
0000$-$263 & 3.3901 & $21.40 \pm 0.1$  & $\leq -1.90$      & $\geq -0.29$  & 1 \nl        
0013$-$004 & 1.9730 & $20.70 \pm 0.1$  & $-0.80 \pm 0.1$   & $\leq -0.67$  & 2 \nl
0056$+$014 & 2.7771 & $21.11 \pm 0.07$ & $-1.23^{+0.21}_{-0.38}$  & $-0.12^{+0.20}_{-0.39}$  &  1 \nl
0100$+$130 & 2.3091 & $21.40 \pm 0.05$ & $-1.53 \pm 0.08$  & $-0.18 \pm 0.06$ & 2  \nl
0112$+$029 & 2.4227 & $20.95 \pm 0.1$  & $-1.15 \pm 0.15$  & $-0.50 \pm 0.2$  & 2  \nl
0201$+$365 & 2.462  & $20.38 \pm 0.04$ & $-0.27 \pm 0.05$  & $-0.64 \pm 0.05$ & 3  \nl
0454$+$039 & 0.8596 & $20.76 \pm 0.03$ & $-0.83 \pm 0.09$  & $-0.19 \pm 0.1$  & 4  \nl
0458$-$020 & 2.0395 & $21.65 \pm 0.1$  & $-1.23 \pm 0.14$  & $-0.36 \pm 0.1$  & 2  \nl
0528$-$250 & 2.140  & $20.70$          & $\leq -1.01$      & $\geq -0.24$     & 5  \nl    
0528$-$250 & 2.811  & $21.26$          & $-1.03$           & $-0.32$          & 5  \nl    
0841$+$129 & 2.3745 & $20.95 \pm 0.1$  & $-1.35^{+0.13}_{-0.19}$  & $-0.29^{+0.17}_{-0.26}$  & 1 \nl
0935$+$417 & 1.3726 & $20.40 \pm 0.1$  & $-0.80 \pm 0.14$  & $-0.10 \pm 0.14$ & 6  \nl
1104$-$180 & 1.6616 & $20.8$           & $-0.80 \pm 0.11$  & $-0.49^{+0.16}_{-0.27}$ & 7 \nl
1151$+$068 & 1.7736 & $21.30 \pm 0.1$  & $-1.56 \pm 0.14$  & $-0.08 \pm 0.1$  & 1  \nl
1223$+$178 & 2.4658 & $21.50 \pm 0.1$  & $-1.68 \pm 0.13$  & $-0.14 \pm 0.15$ & 2  \nl
1328$+$307 & 0.6922 & $21.28$          & $-1.21$           & $-0.45$          & 8  \nl     
1331$+$170 & 1.7764 & $21.18 \pm 0.05$ & $-1.53^{+0.18}_{-0.31}$  & $-0.40^{+0.19}_{-0.35}$  & 2  \nl
2230$+$025 & 1.8642 & $20.85 \pm 0.1$  & $-0.56 \pm 0.12$  & $\leq -0.52$     & 2  \nl
\enddata
\tablenotetext{a}{cm$^{-2}$}
\tablerefs{(1) Pettini et al. 1997; (2) Pettini et al. 1994; (3) Prochaska \& Wolfe 1996; 
(4) Steidel et al. 1995; (5) Meyer \& Roth 1990;  (6) Meyer, Lanzetta, \& Wolfe 1995; 
(7) Smette et al. 1995; (8) Meyer \& York 1992  }
\end{deluxetable}
\newpage 
\begin{figure}
\figurenum{1}
\epsscale{1.1}
\plotone{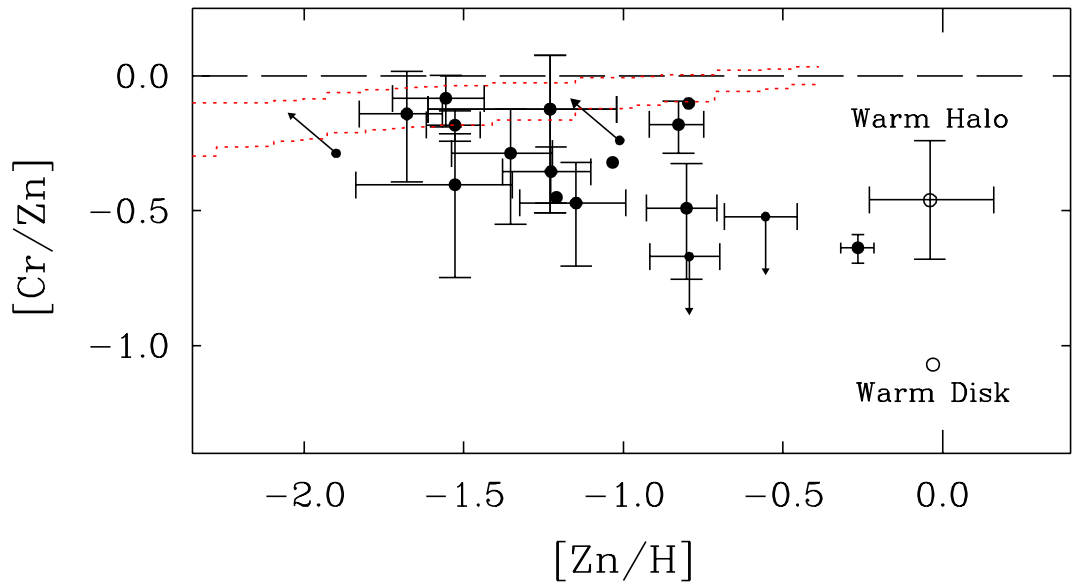}
\vspace{-10cm}
\caption{Cr abundance relative to Zn in 18 damped \lya\ systems (filled
symbols) and in warm interstellar clouds in the disk 
and halo of the Milky Way (open circles, from Savage \& Sembach 1996).
The region within the dotted lines (reproduced from Ryan et al. 1996) 
indicates how the [Cr/Fe] ratio varies in 
Galactic stars in this metallicity regime.} 
\end{figure}

\end{document}